\documentclass{ar-1col}
\usepackage[numbers]{natbib}
\usepackage{amsmath, amsthm, amssymb,slashed,url,tensor}

\makeatletter
\let\oldcitation\citation
\def\citation#1{%
\@for\tmp:=#1\do{%
\global\@namedef{ZZ\tmp}{}%
\oldcitation{\tmp}}}
\let\oldbibitem\bibitem

\renewcommand\bibitem[2][]{%
\expandafter\ifx\csname ZZ#2\endcsname\relax
\color{red}%
\else
\color{black}%
\fi
\oldbibitem[#1]{#2}}%
\makeatother

\jname{Xxxx. Xxx. Xxx. Xxx.}
\jvol{AA}
\jyear{YYYY}
\doi{10.1146/((please add article doi))}

\newcommand{\bs}[1]{\boldsymbol{#1}}
\newcommand{\beq}{\begin{equation}}
\newcommand{\eeq}{\end{equation}}
\newcommand{\la}{\left\langle}
\newcommand{\ra}{\right\rangle}

\begin{document}

\bibliographystyle{ar-style4.bst}

\markboth{P. McClarty}{Topological Magnons}

\title{Topological Magnons: A Review}

\author{Paul McClarty
\affil{Max Planck Institute for the Physics of Complex Systems, N\"{o}thnitzer Strasse 38, 01187 Dresden, Germany.; email: pmcclarty@pks.mpg.de}}

\begin{abstract}
At sufficiently low temperatures magnetic materials often enter a correlated phase hosting collective, coherent magnetic excitations such as magnons or triplons. Drawing on the enormous progress on topological materials of the last few years, recent research has led to new insights into the geometry and topology of these magnetic excitations. Berry phases associated to magnetic dynamics can lead to observable consequences in heat and spin transport while analogues of topological insulators and semimetals can arise within magnon band structures from natural magnetic couplings. Magnetic excitations offer a platform to explore the interplay of magnetic symmetries and topology, to drive topological transitions using magnetic fields. examine the effects of interactions on topological bands and to generate topologically protected spin currents at interfaces. In this review, we survey progress on all these topics, highlighting aspects of topological matter that are unique to magnon systems and the avenues yet to be fully investigated.
\end{abstract}

\begin{keywords}
spin waves, topological matter, quantum magnetism, spintronics
\end{keywords}
\maketitle



\maketitle



\section{INTRODUCTION}

The study of topological magnons springs from the realization that Bloch wavefunctions of quasiparticles in crystalline solids can be associated with nontrivial topological invariants or winding numbers and that these have observable consequences. The pioneering works on topological bands structures came about in the wake of the experimental discovery of integer quantum Hall systems. The quantized Hall conductance present in such systems is related to a topological invariant $-$ the Chern index $-$ in Landau levels. Shortly thereafter Haldane formulated a tight-binding model with bands carrying nonzero Chern number thus making the connection between integer quantum Hall physics and band topology \cite{Haldane1988}. But this model merely scratched the surface and by now the exploration of band topology has been one of the most fruitful endeavours of a generation of physicists \cite{hasan2010,bernevig2013topological,yan2017weyl,armitage2018,burkov2018,chiu2016}. Not only does band topology appear in many different guises but it is also rather common both in naturally occurring insulators and semimetals as well as in engineered band structures of light and sound \cite{ozawa2019,mao2018}.

It may also occur in magnetic excitations where it is providing fresh insights into old physics \cite{Sato2019,malki2020topological,li2020topological,bonbien2021topological}. Spin waves were introduced by Felix Bloch in 1930 to account for the temperature dependence of the magnetization of ferromagnets \cite{Bloch1930}. He showed, within the, then, new quantum theory, that they can be thought of as quantized modes, or magnons, composed of single spin-flip excitations about the ordered magnetic background. Later, these single magnon states were interpreted in terms of semi-classical precessional dynamics and, in parallel, the quantum mechanical description of spin waves was extended to antiferromagnets and to nonlinear effects \cite{HP1940,Herring1951,Dyson1956}. 

Now in the early 2020s, quantum magnetism is a mature field showing no signs of senescence. To the contrary, there is a tremendous amount of activity studying exotic magnetic phenomena especially with strong quantum fluctuations and spin-orbit coupling. Relatively new developments in device fabrication have spawned the active field of magnonics with the goal of tailoring and manipulating spin waves in artificial materials. 

This article reviews the current understanding of magnon band topology. Unless otherwise stated, the word ``magnon" is taken to encompass all types of coherent magnetic excitations $-$ for example, spin waves in ordered magnets, triplon excitations in quantum magnets and exchange coupled excitons. 

We begin, in the next section, by briefly showing how nontrivial topology can arise from the excitations of interacting local moments. We then summarize, some of the diverse topological band structures that can arise in such systems (Sections~\ref{sec:Chern} and \ref{sec:dirac}). Among the thousands of known magnetic materials there are many examples of low dimensional physics where couplings between moments act within chains or layers of spins in three dimensional crystals. Genuinely single layer magnetism is also becoming accessible through the controlled deposition of magnetic ions onto substrates and exfoliation of layers in van der Waals magnets \cite{Burch2018,Gibertini2019,Cai2019}. So magnon band structures and hence magnon topology can be explored in 1D, 2D and 3D. Also, magnetic ground states are generally mutable for example under the influence of an applied magnetic field. This opens up the possibility of exploring field-tuned topological transitions at finite energies. 

In addition, the interplay of magnetic and crystalline symmetries has an important role in enriching band topology (Section~\ref{sec:sym_top}) and magnetic materials offer a platform to study band topology in the presence of interactions. Both the protection of single particle physics from the effects of interactions as well as new physics coming from magnon interactions are discussed in Section~\ref{sec:interactions}.

Having surveyed the theory, we turn to the ways that topology can appear in experimental probes (Section~\ref{sec:volume}). One of the core principles of band topology is the bulk-boundary correspondence that attributes surface states to nontrivial bulk topological invariants. We discuss the open problem of detecting topologically protected magnon surface states in Section~\ref{sec:surface}. Finally, we report on the status of experiments exploring magnon band topology (Section~\ref{sec:materials}).


\section{TOPOLOGICAL MAGNETIC EXCITATIONS: THEORY}

\subsection{Generalities}
\label{sec:SW}

In this section, we briefly outline how spin wave spectra may be computed from a lattice model of localized ordered moments coupled by bilinear interactions. Then, from the general linear spin wave problem, we show how to compute the Berry phase and Berry curvature that play a central role in the study of band topology. We also introduce a topological invariant $-$ the (first) Chern index.

We consider the following general exchange Hamiltonian for localized moments defined on some lattice 
\beq
\mathsf{H} = \frac{1}{2} \sum_{ia,jb; \alpha,\beta} {\rm J}_{iajb}^{\alpha\beta} S_{ia}^{\alpha} S_{jb}^{\beta} 
\eeq
where $ {\rm J}_{iajb}^{\alpha\beta}$ are the exchange couplings between pairs of sites with components $\alpha$, $\beta$.  In solid state magnets, the interactions may originate from exchange or magnetostatic dipolar couplings. We suppose that the moments acquire nonzero expectation values $\la S_{ia}^{\alpha} \ra $, either spontaneously at low temperatures or in a small applied magnetic field, with $i,j$ running over the $N$ magnetic primitive cells and $a,b$ running over the $m$ magnetic sublattices in the magnetic primitive cell. The moments are written in a local quantization frame defined with local $z$ component $\hat{\boldsymbol{z}}_a$ along the ordered moment direction as in $\la S_{ia}^{\alpha} \ra \equiv S \delta^{\alpha z}$ so that the magnetic structure is uniform from site to site in this frame. 

The spins can be bosonized via the Holstein-Primakoff representation of spins of size $S$ \cite{HP1940}
\begin{align}
S^{z} & = S - a^\dagger a \nonumber \\
S^{+} & = \sqrt{2S} \sqrt{1 - \frac{a^\dagger a}{2S}} a = \sqrt{2S} \left( 1- \frac{a^\dagger a}{4S} \right) a + \ldots \nonumber \\
S^{-} & = \sqrt{2S} a^\dagger \sqrt{1 - \frac{a^\dagger a}{2S}} = \sqrt{2S} a^\dagger \left( 1- \frac{a^\dagger a}{4S} \right) + \ldots
\label{eq:HPboson}
\end{align}
where the bosons satisfy the usual commutation relations $[a,a^\dagger]=1$. 

Expanding $\mathsf{H}$, the linear terms vanish around the mean field ground state leaving quadratic terms to order $S$
\beq
\mathsf{H}^{(2)}_{\rm SW} = \frac{S}{2} \sum_{\boldsymbol{k}} \boldsymbol{\Upsilon}^{\dagger}(\boldsymbol{k}) \left( \begin{array}{cc} \boldsymbol{\mathsf{A}}(\boldsymbol{k}) &  \boldsymbol{\mathsf{B}}(\boldsymbol{k}) \\  \boldsymbol{\mathsf{B}}^{\star}(-\boldsymbol{k}) &  \boldsymbol{\mathsf{A}}^{\star}(-\boldsymbol{k})  \end{array} \right) \boldsymbol{\Upsilon}(\boldsymbol{k}) \equiv   \frac{S}{2} \sum_{\boldsymbol{k}}  \boldsymbol{\Upsilon}^{\dagger}(\boldsymbol{k}) \boldsymbol{M}(\boldsymbol{k}) \boldsymbol{\Upsilon}(\boldsymbol{k}) 
\label{eq:HSW}
\eeq
where 
\beq
\boldsymbol{\Upsilon}^{\dagger}(\boldsymbol{k}) = \left( \begin{array}{cccccc} a^{\dagger}_{\boldsymbol{k}1} & \ldots & a^{\dagger}_{\boldsymbol{k}m} & a_{-\boldsymbol{k}1} & \ldots & a_{-\boldsymbol{k}m} \end{array}   \right) 
\eeq
and
\begin{align}
\mathsf{A}_{ab}(\boldsymbol{k}) & ={\rm J}_{ab}^{+-}(\boldsymbol{k}) - \delta_{ab} \sum_{c} {\rm J}_{ac}^{zz}(\boldsymbol{0}) \label{eq:Ablock} \\
\mathsf{B}_{ab}(\boldsymbol{k}) & = \frac{1}{2}\left({\rm J}_{ab}^{xx}(\boldsymbol{k}) - {\rm J}_{ab}^{yy}(\boldsymbol{k}) -i {\rm  J}_{ab}^{xy}(\boldsymbol{k}) - i {\rm J}_{ab}^{yx}(\boldsymbol{k})   \right) \nonumber \\
& =  {\rm J}_{ab}^{--}(\boldsymbol{k}).
\label{eq:Bblock}
\end{align}
One observation to make of Eq.~\ref{eq:HSW} is that it generally does not conserve boson number. In this sense it bears some relation to mean field models of superconductivity. However there are crucial differences. First of all, in order to be stable, the bosonic Hamiltonian must be non-negative. Secondly, the bosonic commutation relations must be preserved under any transformation of the bosons. These commutation relations are $\left[ \Upsilon_a , \Upsilon_b^\dagger \right] = \eta_{ab}$ where $\bs{\eta}={\rm diag}(1,1,\ldots,-1,-1,\ldots)$ with $m$ ones and $m$ minus ones along the diagonal. The diagonalizing transformation must then satisfy 
\begin{align*}
\boldsymbol{U}^\dagger (\boldsymbol{k})  \boldsymbol{M}(\boldsymbol{k}) \boldsymbol{U}(\boldsymbol{k}) & = \boldsymbol{\Lambda}(\boldsymbol{k}) \\
\boldsymbol{U}(\boldsymbol{k}) \boldsymbol{\eta}\boldsymbol{U}^\dagger (\boldsymbol{k}) = \boldsymbol{\eta}
\end{align*}
where $\boldsymbol{\Lambda}(\boldsymbol{k})$ is diagonal. Unlike the analogous transformation for diagonal fermionic Hamiltonians, $\boldsymbol{U}$ is evidently not unitary in general. \footnote{This formulation of linear spin wave theory, while useful to see the connections to electronic band topology, obscures the fact that the spectra thus obtained are identical to those obtained from classical precessional equations of motion (see for example \cite{lu2018magnon}). Indeed, much of the following discussion applies equally to microscopic magnets and to artificial arrays of coupled mesoscopic moments.}

Suppose that the eigenstate for the $n$th band is multiplied by a phase that is a smooth function of momentum $\exp(i \theta(\bs{k}))$. The Berry connection defined by
\beq
A_{\nu}^{(n)}(\bs{k}) \equiv i  \left[  \bs{\eta} \bs{U}^\dagger (\bs{k}) \bs{\eta} \partial_{k_{\nu}} \bs{U}(\bs{k})  \right]_{nn}
\eeq
is of paramount importance to the study of band topology. This transforms as $A_{\nu}^{(n)}(\bs{k})  \rightarrow A_{\nu}^{(n)}(\bs{k}) - \partial_{k_{\nu}} \theta(\bs{k})$ under the phase change which is  reminiscent of a gauge transformation of an electromagnetic vector potential. The simplest gauge invariant quantity is the Berry curvature which, in 2D, is
\beq
B^{(n)}(\bs{k}) = \partial_{k_{x}} A_{y}^{(n)}(\bs{k}) - \partial_{k_{y}} A_{x}^{(n)}(\bs{k}).
\eeq
A simple application of Stokes' theorem tells us that the integral of the Berry curvature over the magnetic Brillouin zone is zero whenever $A_{\nu}^{(n)}(\bs{k})$ is a smooth function across the zone. When this is the case, the band is said to be topologically trivial. There are cases where it is impossible to parameterize the eigenstates over the entire zone with a single gauge choice. In this case, one can typically split up the zone into patches that intersect along some curve such that, in each patch, a smooth gauge choice is possible. Along the intersecting boundary parametrized by $\bs{k}_{\rm bdy}$, two gauge choices can be connected by a phase change $A_{1}^{(m)}(\bs{k}_{\rm bdy}) = A_{2}^{(m)}(\bs{k}_{\rm bdy}) - \partial_{k_{\nu}} \theta(\bs{k}_{\rm bdy})$. In this two patch, two-dimensional case, the integral of the Berry curvature over the Brillouin zone
\beq
C^{(n)} \equiv \frac{1}{2\pi} \int d^2\bs{k} B^{(n)}(\bs{k}) = \oint_{\rm bdy} d\bs{k}\cdot \left( \bs{A}_{1}^{(n)}(\bs{k}) - \bs{A}_{2}^{(n)}(\bs{k})   \right) \in \mathbb{Z}.
\label{eq:chern}
\eeq
This integer is called the (first) Chern number and it is robust to deformations of the band structure that do not close the gap to neighbouring bands.

\subsection{Chern Bands}
\label{sec:Chern}

\begin{figure}[h]
\includegraphics[width=\linewidth]{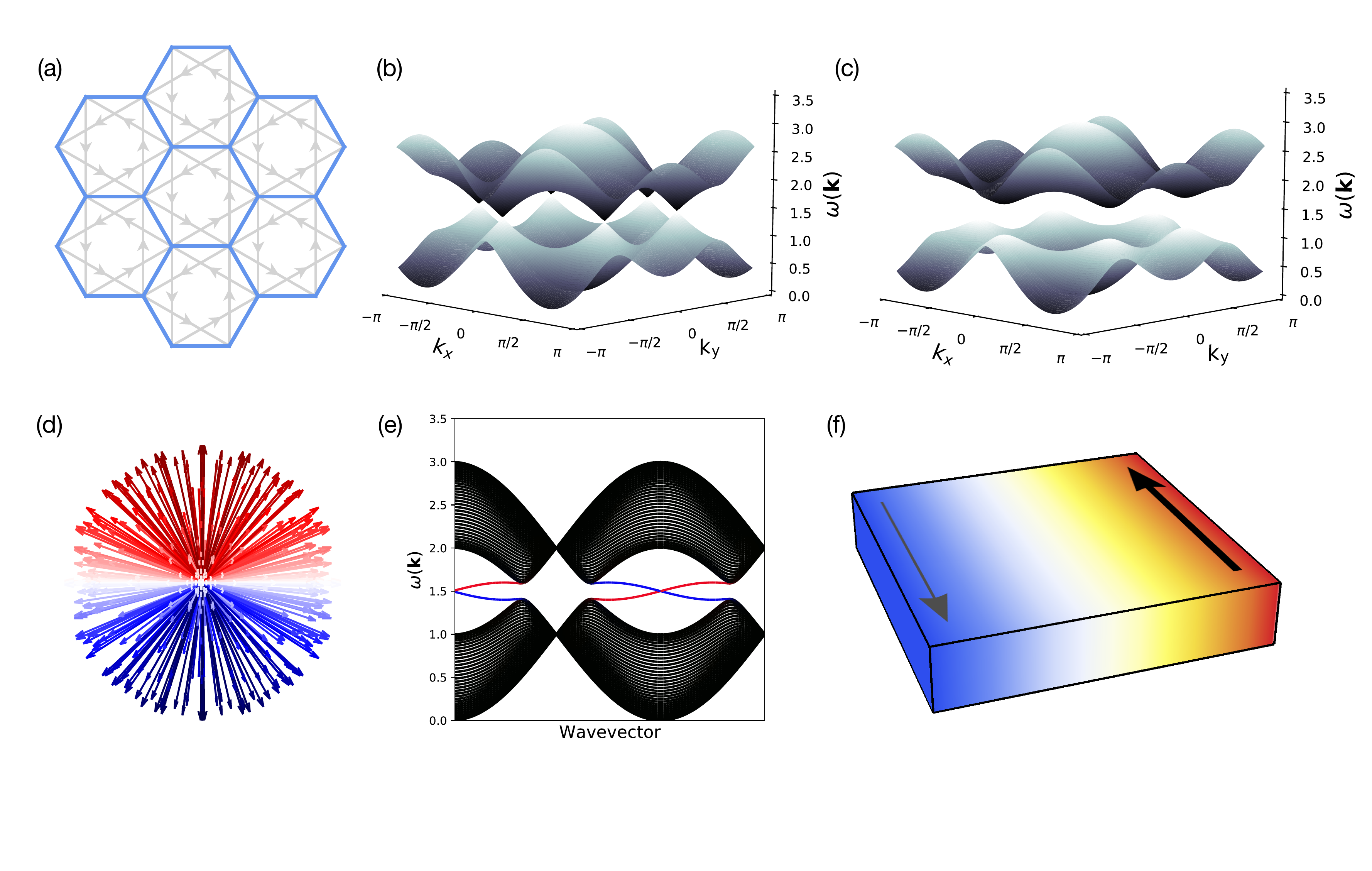}
\caption{Chern magnon bands may exist in 2D magnets with broken effective time reversal symmetry. Panel (a) represents the honeycomb lattice model of Eq.~\ref{eq:honeycombmodel} with Heisenberg exchange acting between nearest neighbours and second nearest neighbour DMI indicated by the grey bonds with the bond orientation indicated and the $\bs{D}$ vector out of plane. When $\vert \bs{D}\vert =0$ there are Dirac points at the zone corners (b) that are gapped out when the DMI is switched on (c). The Berry curvature at each point in the zone for one of the bands is plotted from a common origin in (d) showing that it covers the sphere implying a nonzero Chern number. Panel (e) shows the spectrum of the model on a slab geometry. Modes that live in the bulk are rendered in black and those on opposite boundaries in blue and red. The plot shows that there are chiral modes in the bulk band gap. Panel (f) is a schematic of the thermal Hall effect. The colours represent the thermal gradient across a sample and the arrows indicate the surface magnon currents. The surface current is greater where the magnons are more thermally populated. So although the surface currents are counter-propagating at opposite edges the thermal gradient leads to a net transverse heat current.}
\label{fig:chern}
\end{figure}

We now take a simple example to illustrate the emergence of magnon Chern bands from a microscopic model. The model is one of interacting magnetic moments arranged on a honeycomb lattice (Figure~\ref{fig:chern}(a)).  The moments, of length $S$, are coupled through nearest neighbour ferromagnetic Heisenberg exchange and a second nearest neighbour Dzyaloshinskii-Moriya interaction (DMI) with $\bs{D}$ vector perpendicular to the plane of the lattice \cite{owerre2016first}. The latter coupling is allowed by symmetry because the midpoints of second neighbour bonds are not inversion symmetric. This, coupled to the fact that the DMI tends to appear to leading order in the spin-orbit coupling, makes this a fairly natural model of magnetism on the honeycomb lattice. The Hamiltonian is then
\beq
H = - \vert J \vert \sum_{\la i,j \ra_1} \bs{S}_{i} \cdot \bs{S}_{j} + D \sum_{\la i,j \ra_2} \hat{\bs{z}}\cdot \left( \bs{S}_{i} \times \bs{S}_{j} \right) - \sum_i \bs{h}\cdot \bs{S}_{i}
\label{eq:honeycombmodel}
\eeq
where we have also included a Zeeman term that tunes the moment polarization direction. The collinear ferromagnetic mean field ground state is stable to the presence of small $D$.

The spin wave Hamiltonian is $JS(a^\dagger_i a_j + {\rm h.c.})$ between nearest neighbours while the second neighbour bonds have imaginary coefficient $iDS \left( a^\dagger_i a_j - {\rm h.c.}\right)$. Fourier transforming leads to
\beq
H_{\rm magnon} = \sum_{\bs{k}} \left( a^\dagger_{\bs{k}A} \hspace{0.2cm} a^\dagger_{\bs{k}B}  \right) \bs{M}(\bs{k})  \left( \begin{array}{c} a_{\bs{k}A} \\ a_{\bs{k}B} \end{array} \right)
\label{eq:dvech}
\eeq 
with $ \bs{M}(\bs{k}) = d^{(0)}\bs{1} +  \bs{d}_{\bs{k}}\cdot \bs{\sigma}$ and 
\begin{align}
d_{\bs{k}}^{(0)} & = 3JS + h/2 \\
d_{\bs{k}}^{(1)} & =  JS \sum_{\mu =1,2,3} \cos\left( \bs{k}\cdot \bs{r}^{\mu}_{1} \right) \\
d_{\bs{k}}^{(2)} & =  -JS \sum_{\mu =1,2,3} \sin\left( \bs{k}\cdot \bs{r}^{\mu}_{1} \right) \\
d_{\bs{k}}^{(3)} & = DS \cos\theta  \sum_{\mu =1,2,3} \sin\left( i\bs{k}\cdot \bs{r}^{\mu}_{2} \right)
\label{eq:dvector}
\end{align}
where $\bs{r}_{1/2}^\mu$ are first and second nearest neighbour vectors shown in Fig.~\ref{fig:chern}(a), $\bs{\sigma}=(\sigma_x, \sigma_y, \sigma_z)$ are Pauli matrices and $\theta$ is the polar angle of the moments from $\hat{\bs{z}}$. The Hamiltonian is independent of the azimuthal angle $\phi$ because the exchange Hamiltonian has a global $U(1)$ symmetry. This tight-binding model for magnons is, up to a constant energy shift, identical to the famous Haldane honeycomb model \cite{Haldane1988} first proposed as an example of a tight-binding model with topologically nontrivial bands.

The band dispersions of the model are plotted in Figure~\ref{fig:chern}. There are two bands and the lower energy band has a quadratically dispersing Goldstone mode when the applied field is zero that is lifted to finite energy in a field. When either the DMI vanishes or when the magnetic moments are rotated into the honeycomb plane ($\theta=\pi/2$), there are band degeneracies called Dirac points in the magnon spectrum at finite energy (Figure~\ref{fig:chern}(b)). Otherwise, there is an energy gap between the two magnon bands (Figure~\ref{fig:chern}(c)). When this gap is nonvanishing, the Chern numbers (Eq.~\ref{eq:chern}) of the two bands are $+1$ and $-1$ $-$ the absolute sign of each band depending on the sign of $D$. The Chern number turns out to have a simple geometrical interpretation as a winding number for two band models (Figure~\ref{fig:chern}(d)): it counts the number of times unit vector $\hat{\bs{d}}_{\bs{k}}$ covers a spherical surface as the Brillouin zone is covered (Figure~\ref{fig:chern}(d)). 

So far, the discussion seems academic. But, the Berry curvature and Chern number both have direct physical consequences. Suppose, we introduce a physical boundary into the otherwise infinite system. Because the outside is topologically trivial and the inside has topologically nontrivial bands ($C=\pm 1$), the two bands are forced to cross at the boundary. This means that the magnon model  has states living at the boundary so localized in real space and constrained to lie between the two bulk bands (Figure~\ref{fig:chern}(e)). This is an instance of the bulk-boundary correspondence.

The example of the honeycomb ferromagnet with DMI served to illustrate several important features of topological magnon models. The ground state is topologically trivial yet there are topological modes at finite frequency with protected finite energy surface states. The electronic analogue of this model is well-known to have an integer quantum Hall effect $-$ a quantized Hall conductivity when the Fermi energy is in the band gap. For topological magnon systems, owing to the bosonic statistics of the modes there is no such quantization. However, there is a thermal Hall effect associated to the Berry curvature (Section~\ref{sec:volume}). 

A wide variety of exchange models with magnon Chern bands are known beyond the example given above \cite{owerre2016first}. They are necessarily two-dimensional magnets but, beyond this, they have little in common. In the honeycomb example, DMI is responsible for the appearance of Chern bands. The same is true of many other models on different lattices with different ground states including ferromagnets, antiferromagnets, and skyrmion textures as well as in artificial magnonic arrays  \cite{zhang2013,mook2014edge,seshadri2018,laurell2018kagome,laurell2017,cao2015lieb,LeeKiHoon2018,mook2020interactionstabilized,malki2019magnonics,bhomick2020,iacocca2017,Rold_n_Molina_2016,diaz2019skyrmion,diaz2020edge,ymn6sn6}. Chern magnon bands can arise from other types of coupling including Kitaev and off-diagonal symmetric exchange  \cite{mcclarty2018, joshi2018}, long-range magnetostatic dipolar couplings \cite{shindou2013topological,shindou2013edge}, nearest neighbour exchange of dipolar character \cite{honeycombpseudodipolar} and even simple XXZ couplings for noncoplanar spin textures  \cite{kim2019}.

We have focussed our attention on models with magnetic ground states that break spin rotational symmetry and physical time reversal. Magnetic insulators overwhelmingly tend to have such ground states but there are important and interesting exceptions. Among these are dimerized quantum magnets so-called because the lattice structure favours the formation of singlets from pairs of localized spin one-half moments. Here too there are coherent magnetic quasiparticle excitations called triplons from which Chern bands may arise. One such example is the Shastry-Sutherland model with DMI in a small magnetic field that exhibits topological transitions  \cite{romhanyi2015hall,mcclarty2017topological,malki2017}. Chern bands have been found also in other models of quantum magnets such as a triangular lattice of trimers \cite{romhanyi2019} and a kaleidoscope of different Chern triplon band phases have been found in Kitaev-Heisenberg honeycomb triplons \cite{anisimov2019}. 

In electronic systems, the first known Chern bands were Landau levels in two dimensional electron gases in semiconductors from which the integer quantum Hall effect arises. Landau levels are formed from (nearly) free electrons in static homogeneous magnetic fields. But effective magnetic fields can arise in other ways for example in electrons propagating through noncollinear spin textures or through mechanically strained backgrounds. In this spirit, two classes of proposal have been put forward to engineer Landau levels of magnons. The first is to apply strain to a two-dimensional magnet. In the vicinity of quadratically dispersing points in the spectrum, the strain enters through a minimal coupling to a pseudo vector potential $\bs{A}$: $(1/2m)\left( \bs{p} + e\bs{A}/c \right)^2$ where for appropriate strain fields $\bs{A}$ can correspond to a uniform effective magnetic field. For example, Ref.~\cite{nayga2019} shows that triaxial strain in a honeycomb Heisenberg antiferromagnet (exchange $J$) in the vicinity of the quadratically dispersing $K$ points at the top of the single magnon spectrum leads to magnon Landau levels with estimated splitting $10^{-3}J$. Another route to obtaining magnon Landau levels is to exploit the Aharonov-Casher effect whereby magnetic dipoles pick up a Berry phase $\phi_{ij}$ in an electric field gradient along a path from $\bs{r}_i$ to $\bs{r}_j$. Alongside other electric field effects in magnetic insulators there is such a phase that, for moments oriented along $\hat{\bs{z}}$, is
\beq
\phi_{ij} = \frac{g \mu_{\rm B}}{\hbar c^2} \int_{\bs{r}_i}^{\bs{r}_j} d\bs{r} \bs{E}(\bs{r})\times \hat{\bs{z}}
\eeq
where $\bs{E}$ is the electric field and $g$ is the g factor. Landau level splittings of the order of $1\mu$eV have been estimated for this mechanism \cite{nakata2017qhe,nakata2017magnonic}.

\subsection{Dirac Points, Nodal Lines and Weyl Points}
\label{sec:dirac}

We now turn our attention to degeneracies in magnon spectra including magnon band touching points and lines. We have already seen that the honeycomb Heisenberg ferromagnet has a magnon band crossing at the zone corners with linear dispersion $\omega(\bs{k}) = v\vert \bs{k}\vert $ in the vicinity of the touching points. By expanding about the degenerate point at $K$ to linear order in the momentum, the resulting effective Hamiltonian is of the form $H_{\rm eff}(\bs{k}) = k_x \sigma_x + k_y \sigma_y$ which is the massless two-dimensional Dirac equation, where the spinor index corresponds to a sublattice label, so the touching points are called Dirac points. 
The Dirac points are protected here by effective time reversal and inversion symmetries and are therefore gapped out by arbitrarily weak symmetry-breaking perturbations.

It is possible to devise realistic models for several different lattices and magnetic structures where Dirac points appear in magnon spectra $-$ in other words where the protecting symmetries are present at least at the linear spin wave level \cite{fransson2016,boyko2018,BoykoDirac2020,zhai2020,Okuma}. This includes the generalization of Dirac crossings to three dimensions where they are four-fold degenerate \cite{LiDirac2017}. The presence of touching points in reasonable models has been a motivating factor in exploring some of their detailed consequences such as the effects of interactions, strain and electric field gradients \cite{fransson2016,shivam2017,pershogub2018dirac,ferreiros2018,kumar2020,hwang2020}.

By stacking honeycomb layers and coupling them while preserving effective time reversal and inversion symmetries, the Dirac point becomes an extended object in momentum space $-$ a nodal line that has an associated winding number. By now several models have been devised that have nodal lines in their spin wave spectra $-$ whether running through the entire zone or forming closed loops within the zone \cite{LiDirac2017,pershogub2018dirac,mook2017nodal,OwerreNodalLoops,Bao2018,yao20183dafm,yuan2020dirac,elliot2020visualization,liu2020dipolar,li2018spinone}. 

Another type of band degeneracy in three dimensions are linear touching points of pairs of bands called Weyl points. In the simple isotropic case, the states close to the Weyl points are spinor solutions of $-i\partial_t \psi_{\pm} = \pm H_{\rm eff}(\bs{k})\psi_{\pm}$ where  $H_{\rm eff} = v \bs{k} \cdot \bs{\sigma}$ where the $v$ is the magnon velocity and $\bs{k}$ is a three-component momentum measured relative to the Weyl points. The sign indicates whether the isospin polarization of the magnons is parallel or antiparallel to the momentum analogous to the helicity of massless relativistic fermions. 

Weyl points carry a nonzero Chern number computed over a surface enclosing the touching point. This is equivalent to saying that they are quantized charges of Berry flux. For this reason they are robust features of the band structure protected by the separation in momentum space from counterparts with opposite Chern number. Just as with Chern bands in 3D, there is a bulk-boundary correspondence. Projecting oppositely ``charged" Weyl points onto a surface, the surface states are chiral modes extending along arcs connecting the projected points.

\begin{figure}[!htb]
\includegraphics[width=\linewidth]{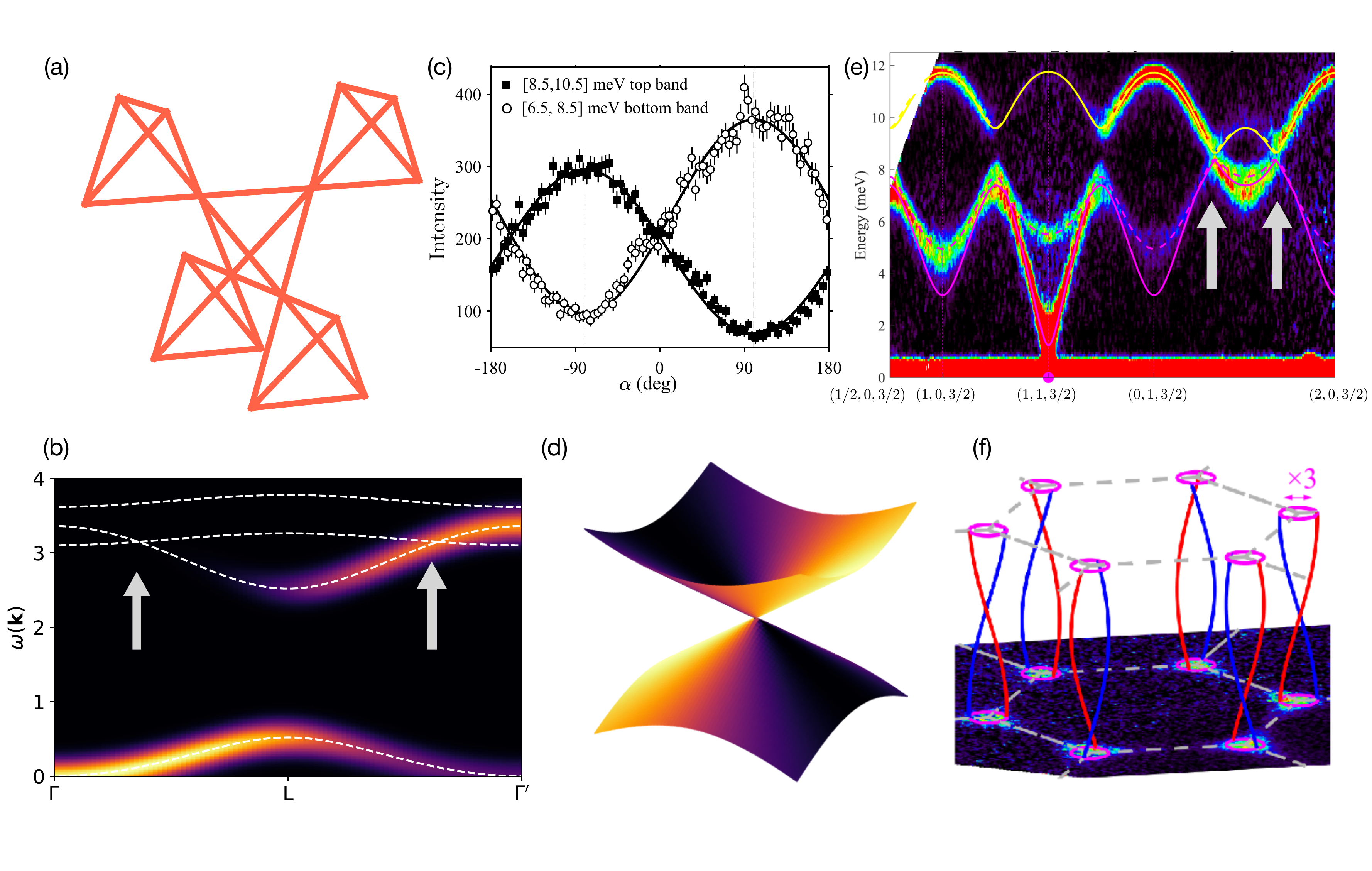}
\caption{The pyrochlore lattice (a) is a network of corner-sharing tetrahedra and a ferromagnetic Heisenberg model on this lattice with nearest neighbour DMI has Weyl points in the magnon spectrum between the middle two bands in order of energy \cite{zhang2013,mook2014edge,seshadri2018}. Panel (b) shows the dynamical structure factor along a momentum-energy cut for this model when the moments are aligned in the $[111]$ direction. Dispersions are indicated by dashed lines revealing two Weyl points. Around linear touching points of magnons, in any context, the neutron scattering intensity varies as $\cos(\alpha-\alpha_0)$  on a loop around the touching point in, say, the upper band and in anti-phase for the lower band where $\alpha$ parametrizes the angle around the loop in momentum space (with some offset angle $\alpha_0$). This feature is illustrated in (d) and experimental neutron scattering intensity around nodal lines in CoTiO$_3$ is shown in (c) \cite{elliot2020visualization}. Inelastic neutron scattering data on CoTiO$_3$ in (e) shows the dispersions along an energy-momentum cut showing touching points marked with arrows. Panel (f) is a schematic of the double helix nodal lines that wind in the out-of-plane direction and that arise in the model parameterizing the neutron data taken on CoTiO$_3$ \cite{yuan2020dirac,elliot2020visualization}. Panels (c), (e), (f) reproduced from Ref.~\cite{elliot2020visualization}.
}
\label{fig:DiracWeyl}
\end{figure}

Exactly this physics may arise in magnon systems. As a concrete example, we mention a 3D analogue of the kagome ferromagnet with DMI that is known to have Chern bands \cite{zhang2013,mook2014edge,seshadri2018}. This analogue is defined on the pyrochlore lattice which is a lattice of corner-sharing tetrahedra (Figure~\ref{fig:DiracWeyl}(a)). A model with $1$st and $2$nd nearest neighbour ferromagnetic Heisenberg and DMI has a pair of Weyl points that originate from a quadratic touching point that is present in the Heisenberg case \cite{mook2016weyl,su2017weylpyrochlore}. The locations of the Weyl points can be tuned by an external magnetic field $\bs{h}$ that rotates the net magnetization. In fact, for high symmetry directions, the Weyl points are located at $\pm \bs{K}$ where $\bs{K} \parallel \bs{h}$. Figure~\ref{fig:DiracWeyl} shows the dynamical structure factor for this model with $[111]$ polarized moments along the $\Gamma L \Gamma' $ cut through the Brillouin zone. Similarly to the case of Chern bands, Weyl points can arise on many different lattices \cite{zhang2020weyl,zyuzin2018weyl}, with different exchange couplings such as the breathing pyrochlore with DMI \cite{Li2016}, a stacked kagome lattice \cite{owerre2018weylkagome}, a stacked honeycomb lattice \cite{su2017weylhoneycomb,lijiang2017weylhoneycomb}, in pyrochlore antiferromagnets \cite{li2018spinone,jian2018weyl} and quantum magnets \cite{bhowmick2020}.

\subsection{Interplay of Symmetry and Topology}
\label{sec:sym_top}

So far we have explicitly introduced one topological invariant that may arise in magnon band structures: the first Chern number. But this is merely the tip of the iceberg. The key to finding more types of nontrivial band topology and to classify them is to organize Hamiltonians according to their dimensionality and their symmetries. For bosonic systems, time reversal symmetry is central to the classification of band topology in the absence of lattice symmetries \cite{chiu2016,xuthreefold}. Time reversal is associated with an anti-unitary operator $T$ that leaves the spin wave Hamiltonian invariant and that squares to plus or minus one.

In the absence of time reversal and in two dimensions, the Chern number is the topological index that distinguishes equivalence classes of gapped Hamiltonians. Since physical time reversal is broken by spontaneous magnetic order or in an applied magnetic field, the conditions for Chern magnon bands in 2D and Weyl points in 3D to exist are easily accommodated by magnetic systems.

But {\it effective} time reversal can be present in magnon Hamiltonians even when {\it physical} time reversal is broken. For example the honeycomb Heisenberg ferromagnet introduced in Section~\ref{sec:Chern} has the property that $\bs{M}^*(\bs{k}) = \bs{M}(-\bs{k})$ implying that the antiunitary complex conjugation operator is an effective time reversal operator that leaves the Hamiltonian invariant. This symmetry (together with inversion or $C_2$) preserves the gapless Dirac points.

Effective time reversal symmetry, with $T^2=-1$, can protect gapped band topology in two and three dimensions and is associated with a two-valued, or $\mathbb{Z}_2$ topological index. In magnon bands, it is possible to engineer models with such a symmetry \cite{Kondo2019a,Kondo2019b}. The main idea explored so far is to find a model where the magnon bands have nonzero Chern number and to couple a time-reversed copy so that the whole system has effective time reversal symmetry. This is naturally accomplished using magnetic bilayers where each layer is an extended two-dimensional lattice model carrying magnons with nonzero Chern number as in the examples given above. One concrete example is a kagome bilayer model that also realizes a magnon analogue of the spin Hall insulator first described by Kane and Mele \cite{KaneMele2005a,KaneMele2005b}. Similar physics has been shown to arise in dimer magnets on bilayers \cite{Joshi2019,thomasen2020novel}.

So far we have discussed time reversal symmetry in isolation but magnon bands are usually highly constrained by sublattice \cite{kovalev2018chiral} and crystalline symmetries as well and this can lead to a proliferation of different types of topological bands as is known from work on electronic systems. Works highlighting the effect of magnetic space group symmetries on bulk band degeneracies include \cite{choi2019} and \cite{corticelli2021} $-$ the latter highlighting the role of decoupled spin rotation symmetry in protecting band degeneracies.  

The classification of possible topological magnon systems according to their symmetry is a topic of on-going research. As we saw, the diagonalizing transformation for free bosons, $\boldsymbol{U}$, is not generally unitary. But it can be related to a unitary problem \cite{colpa1978} thus making the connection with existing fermionic classification schemes \cite{lu2018magnon}. Another route is to recognize that the spin wave spectrum can be obtained by diagonalizing the non-Hermitian matrix $\boldsymbol{\eta M}(\boldsymbol{k})$ that reduces the problem of classification to that of particular symmetry classes of non-Hermitian Hamiltonians \cite{Kondo2020}. 

\subsection{Effects of Magnon Interactions}
\label{sec:interactions}

In electronic systems, interaction effects in the vicinity of a Fermi surface fall off rapidly as the possible scattering phase space drops away as $(\epsilon-\epsilon_{\rm F})^2$ from Fermi energy $\epsilon_{\rm F}$. Magnons, being finite energy excitations, often have much less restrictive kinematic conditions on their scattering processes and their effects have been explored in quantum magnets.

Let us first establish as directly as possible some simple facts about spin wave theory beyond quadratic order. Previously we expressed the ordered moment in terms of Holstein-Primakoff bosons (Eq.~\ref{eq:HPboson}) and linear spin wave theory about a magnetically ordered state corresponds to the $O(S)$ terms in the expansion. The effects of interactions on the single magnon states can be computed systematically in perturbation theory for example including terms of $O(S^0)$: the cubic and quartic terms in the spin wave expansion \cite{zhitomirsky2013}
\begin{align}
H_3 & = \frac{1}{\sqrt{N}} \sum_{\bs{k},\bs{k}'} \sum_{abc} [V^{(3)}]^{abc}_{\bs{k},\bs{k}'} a^\dagger_{\bs{k}a}a^\dagger_{\bs{k}'b}a_{\bs{k}+\bs{k}'c} + {\rm h.c.} \nonumber \\
H_4 & = \frac{1}{N} \sum_{\bs{k},\bs{k}'\bs{q}} \sum_{abcd} [V^{(4)}]^{abcd}_{\bs{k},\bs{k}',\bs{q}} a^\dagger_{\bs{k}+\bs{q}a}a^\dagger_{\bs{k}'-\bs{q}b}a_{\bs{k}c}a_{\bs{k}'d} + [V^{(4)}]^{abcd}_{\bs{k},\bs{k}',\bs{q}} a^\dagger_{\bs{k}a}a^\dagger_{\bs{k}'b}a^\dagger_{\bs{q}c}a_{\bs{k}+\bs{k}'+\bs{q}d}  + {\rm h.c.}
\label{eq:interactionterms}
\end{align}
To examine the effects of interactions on the single magnon states one is interested in the retarded Green's function
\beq
\bs{G}^{\rm R}(\bs{k},\omega) = \left[ \left( \omega + i0^+ \right) - S \bs{M}(\bs{k}) + \bs{\Sigma}^{\rm R}(\bs{k},\omega) \right]^{-1}.
\eeq
When interactions are switched off, this is the single magnon propagator depending only on the linear spin wave Hamiltonian $\bs{M}(\bs{k})$. Interaction effects are contained within the energy and momentum dependent self energy $\bs{\Sigma}^{\rm R}(\bs{k},\omega)$. The real part of the self-energy simply renormalizes the single magnon spectrum. The number non-conserving terms in the spin wave Hamiltonian are responsible for coupling single to multi-magnon states potentially leading to spontaneous magnon decay. The resulting lineshape appears as an imaginary part in the self-energy.  

\begin{figure}[!htb]
\includegraphics[width=\linewidth]{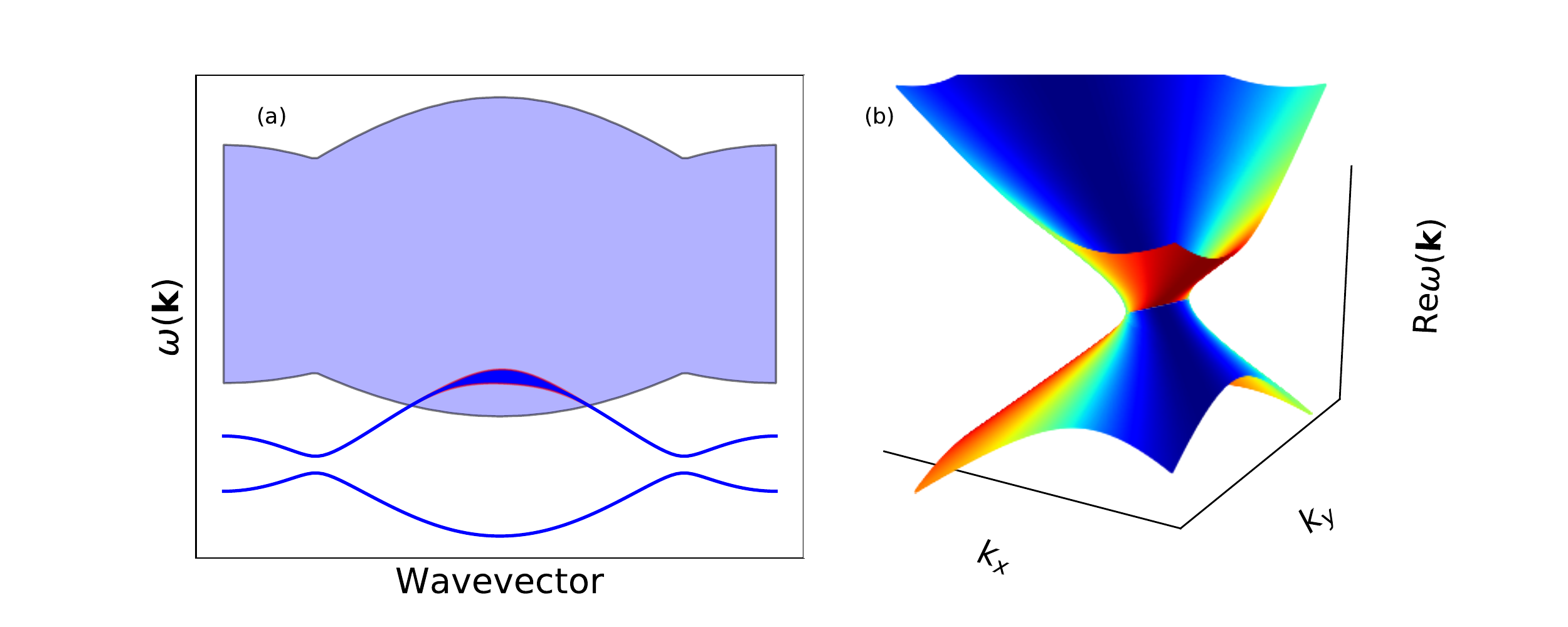}
\caption{Panels illustrating features of interacting magnon systems. (a) Schematic showing the interplay between multi-magnon states and magnon decay taking an example with two dispersive magnon bands with a gap between them. The two-magnon states form a continuum (light blue). Where the single magnon bands overlap with the continuum, decay processes of single magnons into pairs of magnons are kinematically allowed leading to broadening of the single magnon spectrum (indicated). (b) Effect of magnon interactions on a Dirac point in 2D that is immersed in a two-magnon continuum (not shown). As described in the main text, the effective single magnon Hamiltonian is non-Hermitian leading the Dirac point to split into a pair of exceptional points. The dispersions shown are the real part of the spectrum with a nodal line joining the two exceptional points. The colour indicates the imaginary part of the spectrum showing a winding around the entire structure in momentum space that is also in anti-phase in the upper and lower bands and that corresponds to an observable winding of the magnon lineshape. 
}
\label{fig:Interactions}
\end{figure}

The two-magnon states form a continuous band of states in energy-momentum subject to energy-momentum conservation $\left\vert \bs{k}\ra \rightarrow \left\vert \bs{k}-\bs{q}, \bs{q} \ra$ and $\omega(\bs{k}) = \omega(\bs{k}-\bs{q}) + \omega(\bs{q})$. Now the condition for magnon decay has a simple geometrical interpretation (Fig.~\ref{fig:Interactions}(a)): a single magnon can decay when it is both dynamically allowed through the presence of number non-conserving terms in the Hamiltonian and when the single magnon states overlap with the two magnon continuum. 

It follows that when a magnetic field is applied and the energy of the single magnon modes increases, the two-magnon states are lifted at roughly twice the rate. This means that there is a threshold field beyond which single magnon decay is kinematically forbidden. So, at least in principle, the single particle picture and, hence, magnon band topology are robust features of real magnetic materials. The field tunability of magnon damping has been explored non-perturbatively in the field-polarized Kitaev-Heisenberg model \cite{mcclarty2018} showing that topologically protected edge states can be resolved above the threshold field set by linear spin wave theory as expected. Also, at lower fields, one finds that there is significant level repulsion between the two-magnon and single magnon states that is not well captured by interacting spin wave theory from Eq.~\ref{eq:interactionterms} implying that the edge states are more robust than one would expect on the basis of non-linear spin wave theory. 

This favourable situation may not apply to all cases of magnon Chern bands however. For example, in the kagome ferromagnet with Heisenberg exchange and DMI, the fate of the magnon bands in  the presence of magnon interactions has been considered in Ref.~\cite{chernyshev2016damped}. In addition to being responsible for the existence of Chern bands at the level of linear spin wave theory, the DMI also leads to cubic terms and magnon damping. Because the highest energy magnon band is nearly flat, the two-magnon continuum has an unusually high density of states at the zone centre in the band gap separating the two highest energy bands leading to a linewidth of the upper band that is similar to the band gap potentially leading to hybridization of the surface states with bulk states. 

Another case of interest is the problem of linear magnon touching points subject to magnon interactions and where the touching point is immersed within the two magnon continuum. Suppose that the touching point is localized at momentum $\bs{k}_0$ and unrenormalized energy $\omega_0$. It is illuminating to treat $S \bs{M}(\bs{k}) + \bs{\Sigma}^{\rm R}(\bs{k},\omega_0)$ for the two bands in the vicinity of the touching point as an effective Hamiltonian $-$ one that is non-Hermitian since $\bs{\Sigma}^{\rm R}$ is complex. The complex eigenvalues of an entirely generic non-Hermitian perturbation to a Dirac Hamiltonian in 2D correspond to a band structure with a pair of so-called exceptional points where the Hamiltonian has only a single eigenvector. The exceptional points are topologically protected by equal and opposite winding numbers meaning that they cannot be removed unless they annihilate (Figure~\ref{fig:Interactions}) \cite{bergholtz2021}. This non-Hermitian topology has observable features: the single magnon lifetime winds from high to low around the pair of exceptional points and in antiphase between the upper and lower bands \cite{mcclarty2019nonhermitian}.  Interacting magnon systems therefore offer a good platform to study non-Hermitian topology experimentally in quantum magnets.

\section{EXPERIMENT}
\label{sec:expt}

\subsection{Bulk Probes}
\label{sec:volume}

In bulk materials there are well-established techniques for detecting magnons including Brillouin light scattering, inelastic X-ray and neutron scattering (INS). To date, INS has proven one of the most valuable experimental tools in the investigation of topological magnon systems as it provides detailed information about momentum dependent dispersion relations throughout the zone. As we discuss in the next section, this can be used to establish the existence of nontrivial gapped band topology on the basis of a model with couplings fixed by the data.

While establishing gapped magnon topology relies on theoretical input, the presence of nodal lines and points can, at least in principle and often in practice, be observed directly from INS data. In fact, the effect of nontrivial nodal topology on the magnon wavefunctions is directly observable using bulk probes. For linear touching points and lines of magnons in 2D or 3D magnets, the effective Hamiltonian in their vicinity takes a universal form. For Weyl points this is $v \bs{k}\cdot \bs{\sigma}$. This means that the wavefunctions have a pseudospin polarization that is locked to the momentum $\bs{k}$. This pseudospin-momentum locking has direct and universal experimental implications for, as one can show \cite{shivam2017}, the intensity itself winds from high to low when encircling one of the touching points say in the uppermost band. And it winds in antiphase in the lower band. This effect has been observed around the nodal lines in CoTiO$_3$ \cite{elliot2020visualization} (see Figure~\ref{fig:DiracWeyl}).
 
Alongside local probes, transport measurements are extremely useful for characterizing materials. For uncharged particles like magnons, the quantities one can hope to probe via transport measurements are thermal conductivity and, in some instances, spin currents. We begin the discussion with the thermal Hall effect. This is a transverse thermal conductivity $\kappa_{xy}$ in the presence of a temperature gradient applied along $\hat{x}$ in a time reversal broken system analogous to the electrical Hall effect. 

Among other possible origins, a thermal Hall effect can arise from the Berry curvature of magnons \cite{murakami2016thermal}. To see this, we look at the semiclassical equations of motion of magnon wavepackets  \cite{katsura2010theory,matsumoto2011prl,matsumoto2011rotational}
\begin{align}
\dot{\bs{r}} & = \frac{1}{\hbar} \frac{\partial \omega^{(n)}(\bs{k})}{\partial \bs{k}} - \dot{\bs{k}} \times \bs{B}^{(n)}(\bs{k}) \\
\hbar  \dot{\bs{k}} & = - \nabla U(\bs{r})
\end{align}
where the wavepacket is centred at $\bs{r}$ and $\bs{k}$ in phase space. Here the dynamics is driven by the gradient of a confining potential $U(\bs{r})$ at the boundary of the sample. As the momentum centre coordinate drifts as a result of this gradient, an anomalous velocity velocity appears $- \dot{\bs{k}} \times \bs{B}^{(n)}(\bs{k})$ that drives the magnons along the edge. This anomalous term is analogous to a Lorentz force with the Berry curvature in mode $n$, $\bs{B}^{(n)}(\bs{k})$, \cite{dugaev2005} playing the role of an effective magnetic field. Of course, for such a response to arise, the magnon chemical potential must be non-vanishing. In addition, because the transport in a homogenous system will cancel at opposite boundaries, either the system must be inhomogeneously pumped or a thermal gradient must be established (Figure~\ref{fig:chern}(f)). A calculation of $\kappa_{xy}$ in a slab geometry gives
\beq
\kappa_{xy} = \frac{2 k_{\rm B}^2 T}{\hbar V} \sum_{\bs{k}, n} c_2( \rho_{\rm B}(\bs{k}) ) B_z^{(n)}(\bs{k})
\eeq
where $c_2$ is a function of the thermal occupation $ \rho_{\rm B}(\bs{k}) $ of the magnon bands.

The first observation of the magnon thermal Hall effect  was in the insulating pyrochlore ferromagnet Lu$_2$V$_2$O$_7$ \cite{onose2010observation} where it was attributed to the presence of a DMI.  
The measurement of $\kappa_{xy}$ on Lu$_2$V$_2$O$_7$ was followed up with similar results for several of its relatives \cite{ideue2012effect}. Since then a thermal magnon Hall effect has been detected \cite{hirschberger2015,kasahara2017thermal,kasahara2018,hentrich2019} or proposed \cite{laurell2018kagome,cao2015lieb,mcclarty2018,joshi2018,kim2019,romhanyi2015hall,nakata2017qhe,matsumoto2014thermal,mook2014magnon,mook2016dynamics,owerre2016honeycombhall,owerre2017kagomehall,cookmeyer2018,mook2019coplanar,hotta2019} in many magnetic systems.

The existence of a thermal Hall effect is neither a necessary nor a sufficient condition for the existence of magnon band topology. But for Chern bands or Weyl points a thermal Hall effect should be present and is therefore a useful measurement for characterization. However, it is straightforward to find examples where the magnon Berry curvature is nonvanishing but where degeneracies preclude the existence of Chern bands (e.g. Ref.~\cite{cookmeyer2018}). Also in time reversal symmetric topological systems, the thermal Hall should vanish.

The foregoing discussion suggests that a nonvanishing thermal Hall effect should coincide with a transverse spin current wherever a conserved spin density can be identified. Sure enough, this so-called spin Nernst effect can arise from a nonvanishing magnon Berry curvature in a thermal gradient \cite{zyuzin2018weyl,kovalev2016fm}. Instead of a thermal gradient one may instead imbalance the magnon chemical potential with a magnetic field again leading to transverse spin current \cite{fujimoto2009}. One of the more amusing features of the spin Nernst effect is that, whereas the thermal Hall effect relies on time reversal symmetry breaking, the spin Nernst effect does not. This means that there are circumstances where the thermal Hall vanishes and the spin Nernst does not: for example in simple antiferromagnets with doubly degenerate magnon bands where the modes have opposite spin chirality \cite{cheng2016,zyuzin2016}. Experiments sensitive to voltages produced by an inverse spin Hall effect on the layered honeycomb antiferromagnet MnPS$_3$ have revealed signatures of the spin Nernst effect  \cite{shiomi2017}.
 
\subsection{Magnon Surface States}
\label{sec:surface}

One of the robust predictions coming from bulk topology is the existence of protected surface states. Chiral surface states of magnons called Damon Eshbach modes have been known to arise in thin film dipolar ferromagnets for some time \cite{damon1961magnetostatic,eshbach1960,mohseni2019} and, in fact, it has been argued relatively recently that they originate from a proximate topological phase \cite{yamamoto2019}. These modes are routinely measured using Brillouin light scattering \cite{grunberg1977,zhang1986} which is appropriate for macroscopic or mesoscopic spin waves. Perhaps the most promising near-term method to study the chiral surface states of Chern magnon bands is in magnonic crystals. These are periodic arrays of mesoscopic ferromagnetic islands. Each island has a large moment so the dominant coupling is the long-range dipolar interaction. It has been shown that such systems can be engineered to have magnon excitations carrying nonzero Chern number \cite{shindou2013topological,shindou2013edge} and the surface states would be accessible to Brillouin light scattering. 

In quantum magnets, the situation is different because the edge states are confined to within a few Angstroms of the surface, because they typically live at much higher energies than Damon Eshbach modes and because they may not be accessible to the small momentum transfers available to Brillouin light scattering. As we review below there are several materials where there is good reason to believe that there are Chern magnon bands or Weyl points. 

Since neutrons interact weakly with matter,  bulk crystal samples (usually several grams) are required in order to measure the spin waves. Conceivably some progress could be made towards measuring surface states using INS $-$ for example measuring an in-gap magnon density of states in a fine powder sample. Inelastic X-ray scattering is another way of measuring magnon dispersion relations that is relatively surface sensitive. Here, however, the principal difficulty is with energy resolution that restricts possible magnetic materials to those with very large exchange couplings $\sim 0.1$ eV. Surface enhanced Raman scattering or THz absorption spectroscopy with intense sources might be adequate to image surface states in cases where such states exist close to the zone centre. 

In light of the difficulty of producing a large enough signal from a microscopically confined surface state, one proposal is to pump magnons with a laser to obtain an exponentially growing population of magnons \cite{malz2019topological}. Among the couplings between photon mode $\alpha$ and magnons is one of the form $a^\dagger_{\bs{k}}a^\dagger_{-\bs{k}}\alpha + {\rm h.c.}$ so pairs of magnons at finite momentum can be created by irradiation. For small couplings this interaction renormalizes the magnon spectrum but beyond some amplitude threshold there is an instability to magnon production that is cut off by nonlinear effects. This could conceivably boost the thermal Hall current perpendicular to the field gradient \cite{malz2019topological}.  

A further idea is to exploit recent advances in making single 2D magnetic layers on a nonmagnetic substrate. If such a magnetic layer were magnetically ordered one could in principle use an STM tip to detect magnons through the inelastic scattering of the tunnelling electron with magnons. As the tunnelling voltage is swept upwards, the conductance will follow roughly the local magnon density of states. By positioning the tip, first within the 2D bulk and then at the edge one can, in principle, detect the presence of an in-gap density of states originating from the protected surface states \cite{feldmeier2020local}.  

While this may be the most feasible starting point in 2D magnets, the holy grail in the study of magnon Chern bands would be the location specific generation of a surface magnon wavepacket, and a downstream measurement of the spin wave current showing that there are truly chiral surface states. Indeed there are theoretical proposals for building various topological spin wave devices including waveguides, spin wave diodes, beam splitters and interferometers \cite{mook2015waveguide,wang2018magnonics,ruckriegel2018}. The realization of such spintronics devices will require overcoming the various significant challenges discussed above. Much remains to be done and the preceding sections have merely provided a fundamental physics case that topological surface states of magnons can exist in materials as well as giving a characterization of their properties. Fortunately, the first hurdle namely the experimental discovery of topological magnons has been overcome as we discuss in the next section.

\subsection{Materials}
\label{sec:materials}

Experimental work on topological magnons is at a relatively early stage in the sense that the huge diversity of possible magnon topology has scarcely been explored. Even so, there have been a number of significant achievements including the characterization of Chern magnon bands and linear touching points in quantum magnets. We give a few examples of each. 

\subsubsection{Cu[$1,3$-benzenedicarboxylate]}

Cu[$1,3$-bdc] is a metal organic framework with magnetic copper ions on a stacked kagome lattice. The scale of the dominant intra-layer ferromagnet exchange  is low in this material such that $T_c \approx 1.77$K. A coupling between the kagome planes is present bringing about antiferromagnetic interlayer ordering albeit with an estimated exchange less than 1\% of the nearest neighbour exchange so that a weak magnetic field is adequate to polarize all the moments. Inelastic neutron scattering on around $2000$ small crystals with $c$ axis co-aligned, reveals three magnon modes with dispersions consistent with the calculated linear spin wave dispersions for a two-dimensional Heisenberg, DMI model \cite{chisnell2015topological,chisnell2016magnetic}. In particular the highest energy mode is nearly flat with a visible gap between it and the next lowest energy band. On the basis of the model, the Chern numbers are found to be $+1$, $0$ and $-1$ from highest to lowest energy. One therefore expects significant Berry curvature in the lowest energy mode that should lead to a thermal Hall effect. And, indeed, the expected magnetic field dependent $\kappa_{xy}$ has been detected in this material \cite{hirschberger2015}. In Section~\ref{sec:interactions}, we described calculations \cite{chernyshev2016damped} that suggest that interaction effects may be severe in this material. This is not evident from the experiment but in this context it would be interesting to revisit this material to investigate the possible effect of interactions on the topological bands.

\subsubsection{SrCu$_2($BO$_3)_2$}

The material SrCu$_2($BO$_3)_2$ is, to a good approximation, a material realization of the Shastry-Sutherland model that lies within the dimerized phase. At zero field there are three distinct gapped ($\sim 3$ meV), weakly dispersing triplon modes \cite{Gaulin2004} with a bandwidth of about $0.4$ meV. This breaking of the triplon degeneracy is a result of DMI couplings predominantly acting between first and second nearest neighbours. On the basis of this minimal set of couplings, one can show  \cite{romhanyi2015hall} that Chern triplon bands arise in the presence of an applied magnetic field. This scenario was examined experimentally in fields up to $2.8$ T which is adequate to explore the evolution of the triplon modes \cite{mcclarty2017topological}. This study revealed the presence of a new low energy mode that hybridizes with the triplon bands that was interpreted as a bound state of triplons in the singlet sector with the hybridrization originating from the DMI. The data allowed detailed refinement of the exchange parameters in the coupled triplon-bound state model from which the existence of Chern bands and a sequence of field-induced topological transitions could be inferred. In common with other systems with Chern magnon bands, one expects a thermal Hall response in SrCu$_2($BO$_3)_2$ which has, to date, eluded experimental detection \cite{Cairns2019} perhaps because significant thermal population of the triplon bands additionally leads to pronounced broadening of the modes \cite{Zayed2014}.

\subsubsection{Ba$_2$CuSi$_2$O$_6$Cl$_2$}

Another topological triplon system \cite{tanaka2019} is Ba$_2$CuSi$_2$O$_6$Cl$_2$ which consists of stacks of bilayers with Cu dimers connecting the layers and negligible couplings between bilayers. There are two triply degenerate triplon modes with a $\sim 2.5$ meV gap and $1$ meV bandwidth with a gap between the degenerate modes. A fit to the dispersions constrains the exchange leading to a model with a chiral symmetry and a nontrivial winding of the effective $\bs{d}_{\bs{k}}$ vector (cf. Eq.~\ref{eq:dvector}) leading to a 2D analogue of the Su-Schrieffer-Heeger model \cite{sshmodel} and protected surface states.

\subsubsection{Cu$_3$TeO$_6$}

While truly two dimensional magnets with Dirac points in their magnon spectra are currently not known, there are three dimensional magnets with nodal line magnons \cite{Bao2018,yao20183dafm,yuan2020dirac,elliot2020visualization}. One such example is the antiferromagnet Cu$_3$TeO$_6$. The low temperature magnetically ordered phase is collinear with $12$ sites in the primitive magnetic cell. Inelastic neutron scattering studies in this phase \cite{Bao2018,yao20183dafm} reveal a rich magnon spectrum that is consistent with doubly degenerate bands. The dispersion relations inferred from the data as well as the intensities are well captured, up to fine structure including a small spectral gap, by linear spin wave theory on a Heisenberg model.  Features of the linear spin wave theory dispersions include three pairs of linear (Dirac) crossing points at the $P$ points in the Brillouin zone as well as $H$ points where three sets of doubly degenerate bands meet. 

\subsubsection{CoTiO$_3$}

A second well-characterized example of nodal topology is the ABC stacked honeycomb material CoTiO$_3$ \cite{yuan2020dirac,elliot2020visualization}. The cobalt moments in this material are subject to an easy plane anisotropy originating from the spin-orbit and octahedral crystal field and the magnetic structure is collinear ferromagnetism within each honeycomb layer and antiparallel moments between layers. Because the magnetic unit cell has four sites, rather than the two on the honeycomb lattice, there are two nodal lines close to the zone corners. One can show that the Dirac nodal lines survive all symmetry-allowed bilinear couplings including bond-dependent exchange couplings. And, the effect of these couplings is to shift the two nodal lines up and down in energy and to buckle them so that they wind in a double helix pattern through the zone in the $[00L]$ direction. Also, the presence of a spectral gap instead of a Goldstone mode at the propagation vector of the magnetic order implies that such anisotropic exchange terms must be present. Inelastic neutron scattering on this material \cite{yuan2020dirac,elliot2020visualization} shows the presence of nodal lines within resolution but the radius of the helical winding is too small to resolve. 

\subsubsection{Cu$_2$OSeO$_3$}

Weyl points have been inferred from inelastic neutron scattering in the ferrimagnet Cu$_2$OSeO$_3$ where the Weyl points are allowed by the non-centrosymmetric crystal structure \cite{zhang2020weyl} with Cu$^{2+}$ ions on a distorted breathing pyrochlore lattice.

\subsubsection{Two Candidate Materials}

In addition to the examples mentioned above, we highlight two further examples where existing experiments point to interesting magnon band topology. The first such example is the van der Waals magnet CrI$_3$. Bulk crystals are ABC stacked honeycomb  ferromagnets  that magnetically order with moments polarized perpendicular to the honeycomb planes. Inelastic neutron scattering in the ordered phase reveals a pair of spin wave branches with an apparent gap at the $K$ points that is argued to arise from next-nearest neighbour DMI as in the detailed example given in Section~\ref{sec:Chern}. In the bulk material there is significant dispersion along $[00L]$ leading to an estimate of the ratio of the interlayer to the intralayer  exchange of about $30\%$ \cite{Chen2018}. This material and its relatives have two main attractive features from the point of view of topological magnons. The first is that they can be prepared as truly 2D magnets on a substrate \cite{Burch2018,Gibertini2019,Cai2019} suggesting that Chern magnons could be realized in single layer systems. The second is that the single ion anisotropy is weak meaning that in-plane polarization in an applied magnetic field could close the measured gap in CrI$_3$ leading to nodal lines in the bulk material.

The second material is Lu$_2$V$_2$O$_7$, introduced in Section~\ref{sec:volume} in the context of the magnon thermal Hall effect. Inelastic neutron scattering reveals magnons with low energy quadratic dispersion and high energy nearly flat bands consistent with a dominant ferromagnetic Heisenberg exchange with subleading DMI \cite{mena2014}. While there is some small discrepancy between DMI extracted from different experiments, the model robustly predicts that Weyl points should be present \cite{mook2016weyl} at high energies $\sim 4$ meV close to the zone centre that have yet to be detected directly.

\section{OUTLOOK}

In the foregoing survey, we have seen ways in which band topology has contributed to our understanding of magnetic excitations through new insights into magnon transport and scattering, magnon decay, and the presence of surface modes. These insights have stimulated new experiments and suggest further promising experimental avenues.

We can look forward to progress at many levels. One of the most outstanding problems is to detect topological surface states of magnons for the first time in quantum materials or magnonic crystals. This on its own could be expected to provide new insights into surface states of magnets including those not of topological origin and, more speculatively, to progress in manipulating microscopic spin currents. This direction looks doubly promising in the light of the growing importance of single layer magnetism and the engineering of magnonic crystals \cite{shindou2013topological,pirmoradian2018}. Theoretical work suggests that Chern magnon bands should be relatively common and more examples are likely to emerge but also the diversity of bulk nodal points, lines and surfaces reinforced by magnetic and crystalline symmetries are all open to new experimental work. Research into magnon transport is proceeding swiftly with the recent discovery of interesting thermal Hall anomalies in $\alpha$-RuCl$_3$  \cite{kasahara2017thermal,kasahara2018,hentrich2019} and the recent spin Nernst experiments \cite{shiomi2017}. These techniques will hopefully become more standard among the repertoire of experimentalists for the light that they shed on Berry curvature. Other new and promising directions include the topology of magnetic excitons and bound states \cite{elliot2020visualization,qin2017,qin2018}, itinerant magnetism, magnon-phonon coupling \cite{park2019,thingstad2019}, higher order topology \cite{park2021hinge,mook2020chiral}, laser control of magnon topology \cite{owerre2017floquet,nakata2019} and other out-of-equilibrium phenomena \cite{kovalev2017pumping}.

\section{DISCLOSURE STATEMENT}
The author is not aware of any affiliations, memberships, funding, or financial holdings that might be perceived as affecting the objectivity of this review.

\section{ACKNOWLEDGMENTS}

My own work on topological magnons has been done in collaboration with Radu Coldea, Alberto Corticelli, Xiaoyu Dong, Miska Elliot, Matthias Gohlke, Tatiana Guidi, Roger Johnson, Frank Kruger, Pascal Manuel, Roderich Moessner, Stuart Parker, Tony Parker, Karlo Penc,  Jeff Rau, Saumya Shivam and Helen Walker. I thank Radu Coldea, David Luitz and Jeff Rau for comments on an earlier version of the manuscript.

\bibliographystyle{unsrt}
\bibliography{references}




\end{document}